\documentclass[proceedings, preprint]{rmaa}

\usepackage{paralist}

\usepackage{psfrag,color}

\SetYear{2017}
\SetConfTitle{Fourth Workshop on Robotic Autonomous Observatories}

\title{On the search of the ``elusive'' Intermediate Mass Black-Holes} 

\author{
  M. D. Caballero-Garc\'{i}a,\altaffilmark{1} 
  S. Fabrika,\altaffilmark{2,3}
  A. J. Castro-Tirado,\altaffilmark{4,5}
  M. Bursa,\altaffilmark{1}
  M. Dov\v{c}iak,\altaffilmark{1}
  A. Castell\'on,\altaffilmark{6}
  and V. Karas\altaffilmark{1}}

\altaffiltext{1}{Astronomical Institute, Academy of Sciences of the Czech 
Republic, Bo\v{c}n\'{\i}~II 1401, CZ-141\,00~Prague, Czech Republic (garcia@asu.cas.cz).}
\altaffiltext{2}{Special Astrophysical Observatory, Nizhnij Arkhyz 369167, Russia.}
\altaffiltext{3}{Kazan Federal University, Kremlevskaya 18, Kazan 420008, Russia.}
\altaffiltext{4}{Instituto de Astrof\'{\i}sica de Andaluc\'{\i}a (IAA-CSIC), P.O. Box 03004, E-18080, Granada, Spain.}
\altaffiltext{5}{Departamento de Ingenier\'ia de Sistemas y Autom\'atica, Escuela de Ingenier\'ia Industrial, Universidad de M\'alaga (Unidad Asociada al CSIC), Spain.}
\altaffiltext{6}{Facultad de Ciencias, Dpto. de \'Algebra, Geometr\'ia y Topolog\'ia, Universidad de M\'alaga. Campus de Teatinos, s/n M\'alaga, Spain.}

\shortauthor{Caballero-Garc\'{i}a et~al.}
\shorttitle{``Elusive'' Intermediate Mass Black-Holes}

\abstract{Ultra-Luminous X-ray sources (ULXs) are accreting black holes for which their X-ray
properties have been seen to be different to the case of stellar-mass black hole
binaries. For most of the cases their intrinsic energy spectra are well described
by a cold accretion disc (thermal) plus a curved high-energy emission components. The
mass of the black hole (BH) derived from the thermal disc component is usually in the
range of 100-$10^{5}$ solar masses, which have led to the idea that this can represent
strong evidence of the Intermediate Mass Black Holes (IMBH), proposed to exist by
theoretical studies but with no firm detection (as a class) so far. Recent theoretical
and observational developments are leading towards the idea that these sources are
instead compact objects accreting at an unusual super-Eddington regime instead. On
the other hand, gravitational waves have been seen to be a useful tool for finding
(some of these) IMBHs. We give a brief overview about the recent advent of the discovery of
gravitational waves and their relationship with these so far elusive IMBHs.
}

\resumen{Los agujeros negros de masa intermedia (IMBH) son una hip\'otesis a\'un no confirmada
con total seguridad (como clase de objetos) a partir de evidencias observacionales directas. Su masa
ser\'{\i}a mayor que la de los agujeros negros de tipo estelar y menor que la de los monstruosos
agujeros negros supermasivos. Unos objetos candidatos a rellenar este hueco lo constituyen las fuentes
de rayos X ultra-luminosas (ULX) en cuyas emisiones se observan propiedades diferentes al caso de los
agujeros negros de masa estelar en sistemas binarios. Sin embargo, desarrollos te\'oricos y observacionales
recientes conducen a la idea de que estas fuentes son, en su lugar, objetos compactos de masa estelar
que acretan en un r\'egimen inusual de super-Eddington. Por otro lado, las ondas gravitacionales se han
visto como una herramienta \'util para encontrar los IMBH. En este art\'{\i}culo damos una breve descripci\'on
sobre el descubrimiento de las ondas gravitacionales y su relaci\'on con estos agujeros negros de masa 
intermedia, hasta el momento esquivos.
}

\addkeyword{Black hole physics}
\addkeyword{Accretion, accretion-discs}
\addkeyword{X-rays: general}
\addkeyword{Galaxies: Formation}
\addkeyword{Cosmology: Early Universe}
\addkeyword{gravitational waves}

\begin{document}
\maketitle

\section{Introduction}
\label{KK}

Ultra-Luminous X-ray sources (ULXs) are point-like, off-nuclear, extra-galactic sources, with observed X-ray luminosities
(${\rm L}_{\rm X}{\ge}10^{39}\,{\rm erg}\,{\rm s}^{-1}$) higher than the Eddington luminosity for a 
stellar-mass black-hole 
(${\rm L}_{\rm X}{\approx}10^{38}\,{\rm erg}\,{\rm s}^{-1}$). The true nature of these objects is still debated
(Feng \& Soria, 2011; Fender \& Belloni, 2012) as there is still no unambiguous estimate for the mass of the compact object hosted in these systems.

Most ULXs are thought to be powered by super-Eddington accretion 
onto a stellar mass black hole which can be accomplished (i) by powering strong disc winds \citep{shakura73,lipunova99}, (ii) by advecting the radiation along with the flow as in radiation pressure dominated 
disc models like Polish doughnuts \citep{abramowicz78,jaroszynski80} and slim discs \citep{abramowicz88}, or (iii) both, advection and outflows \citep{poutanen07,dotan11}. Luminosities up to $10^{41} \, {\rm erg/s}$ can 
therefore still be explained by super-Eddington mass accretion rates onto stellar mass black holes which can have maximum masses up to $\sim 80 \, {\rm M}_{\odot}$. These higher mass black holes can be explained 
by direct collapse of metal poor stars \citep{belczynski10}. The \emph{slim} disc incorporates the effects of advection. This means that with rising mass accretion rate an increasing fraction of photons 
gets trapped in the flow, carried inward, and is partly released at smaller radii. A typical slim disc has higher flux at soft photon energies below the spectral peak and above it in 
comparison to a standard thin disc \citep{straub11,straub14}.

Assuming an isotropic emission, in order to avoid the violation of the Eddington limit, ULXs might be powered by accretion onto Intermediate Mass Black Holes (IMBHs) with masses in the range
$10^{2}-10^{5}\,{\rm M}_{\odot}$ \citep{colbert99,greene07,farrell09}. Recently, some studies have shown evidence of some ULXs being Black Hole Binaries (BHBs, e.g. M~82 X--2; \citealt{kong07,mcaballe13b}). Later
it was shown that M~82 X--2 is a binary but accreting onto a neutron star \citep{bachetti14}. 

In this paper we give an overview of the current models used in the analysis of the X-ray data from ULXs and on the masses derived by using them. Later we justify the use of the slim-disc model for the proper description of these
sources. Finally we discuss on the existence of IMBH and their detections in the electromagnetic spectra and with gravitational waves. 

\subsection{The X-ray properties from Ultra-Luminous X-ray sources}

The spectra of ULXs show a power-law spectral shape in the 3-8\,keV spectral range, together with a high-energy
turn-over at 6-7\,keV \citep{stobbart06,gladstone09,mcaballe10}, and a {\it soft excess} at lower energies (e.g. \citealt{kaaret06}). This {\it soft excess} can be modelled
by emission coming from the inner accretion disc and is characterized by a low inner disc temperature of ${\approx}0.2$\,keV. This
is expected if the black holes in these sources are indeed IMBHs \citep{miller04}. Other explanations for the {\it soft excess}
imply a much smaller mass for the BH in these sources, based on the idea that the accretion in the disc is not intrinsically standard,
in contrast to the majority of BHBs (e.g. see \citealt{kajava09}).

\subsection{The standard accretion disc theory}

The low inner disc temperatures found for some ULXs were initially interpreted as an evidence for the presence of IMBH \citep{miller04}.
In the standard disc-black body model (i.e. Multi-Color Disc Blackbody or MCD; \citealt{makishima86,makishima00}), which is an approximation
of the real standard accretion disc theory, the bolometric luminosity from the accretion disc is calculated as:

\begin{equation} \label{eq1}
 L_{\rm bol}=4{\pi}(R_{\rm in}/{\zeta})^{2}{\sigma}(T_{\rm in}/{\rm f}_{\rm c})^{4}
\end{equation}

Then the mass of the BH derived from the inner disc temperature from the accretion disc is calculated as:

\begin{equation} \label{eq2}
 \frac{{\rm M}_{\rm BH}}{{\rm M}_{\odot}}=20.736{\eta}(\frac{0.41}{\zeta})^{-2}(\frac{1.7{\rm kT}_{\rm in}({\rm keV})}{{\rm f}_{\rm c}})^{-4}
\end{equation}

\noindent Here ${\rm f}_{\rm c}{\approx}1.7$ \citep{shimura95} is the ratio of the color temperature to the effective temperature, or ''spectral hardening
factor'', and ${\zeta}{\approx}0.412$ is a correction factor taking into account the fact that $T_{\rm in}$ occurs at a radius somewhat larger than $R_{\rm in}$
(\citet{kubota98} give ${\zeta}=0.412$) and ${\eta}$ the Eddington ratio (${\rm L}_{\rm bol}={\eta}{\rm L}_{\rm Edd}$). Setting a typical inner disc 
temperature of inner disc temperature of ${\approx}0.2$\,keV (see references above) Eq.~\ref{eq2} gives a typical value 
of ${\rm M}_{\rm BH}=10^{2}-10^{4}\,{\rm M}_{\odot}$ (taking ${\eta}=0.01-1$, i.e. sub-Eddington accretion).

\subsection{The anomalous regime}

However, it has been seen that in a few BHBs (e.g. XTE J1550-564) when they reach a high luminosity level, the L-T relationship departs from what it has been shown above. Then it is 
called that they have entered the anomalous regime \citep{kubota01,kubota04,gierlinski04}.

\begin{figure}[!t]
  \includegraphics[angle=0,clip,width=\columnwidth]{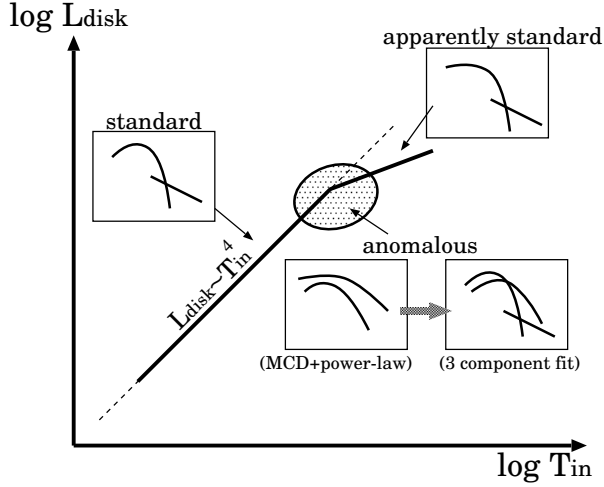}
  \caption{Schematic picture showing the obtained 
spectral regimes on the $L_{\rm disc}$-$T_{\rm in}$ diagram. Thick solid and dashed lines show the source behavior obtained under the MCD
plus power-law fit.The luminosity-temperature relation for super-critically accreting BHs. From \citet{kubota04}. }
  \label{fig1}
\end{figure}

As shown by \citet{kubota01}, these sources generally follow the ${\rm L}_{\rm disc}{\propto}{\rm T}_{\rm max}^4$ relation. When reaching a luminosity of several orders of 
magnitude higher there is a small departure from this: with increasing temperature they seem to be slightly underluminous. This is particularly pronounced 
above ${\rm kT}_{\rm max}=0.9$\,keV where a break in the L-T relation can be seen (Fig.~\ref{fig1}). The data above the break is consistent 
with ${\rm L}_{\rm disc}{\propto}{\rm T}_{\rm max}^{2}$. \citet{kubota04} refer to this as the apparently standard regime, and suggest that it is associated with a transition to a 
slim disc \citep{abramowicz88}. However, \citet{gierlinski04} noted that this should only occur above ${\rm L}_{\rm Edd}$ \citep{abramowicz88,shimura95}, so it seems more likely to 
represent a subtle change in the colour temperature correction factor.

A shift of the microquasar GRO J1655-40 to the right from the $L$--$T$ relation for a 7\,${\rm M}_{\odot}$ BH \citep{orosz97} is probably related to a high spin of the BH there, which results in a higher disc
temperature compared to the Schwarzschild BH.

\subsection{The supercritical regime}

\begin{figure}[!t]
  \includegraphics[angle=0,bb=56 113 509 566,clip,width=\columnwidth]{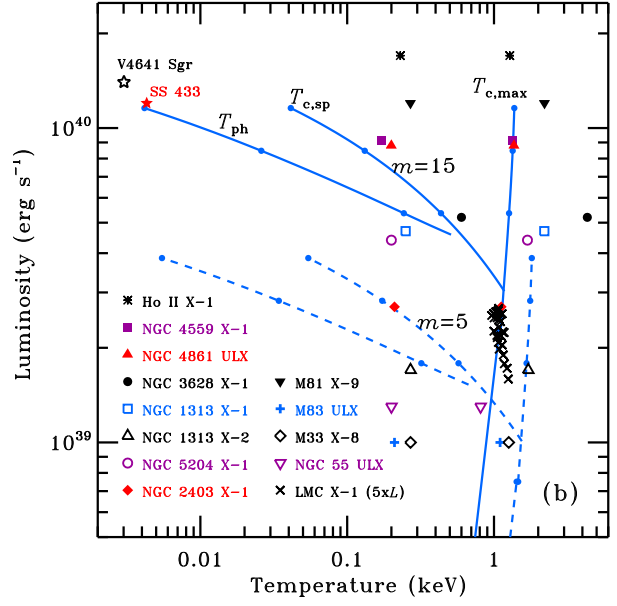}
  \caption{The luminosity-temperature relation for super-critically accreting BHs. From \citet{poutanen07}. }
  \label{fig2}
\end{figure}

A recent study of the spectral variability from a sample of ULXs \citep{kajava09}, has shown that the {\it soft excess} (i.e. the disc component fitted from the X-ray spectra of some canonical ULXs) 
does not follow Eq.~1 but $L_{\rm bol}{\propto}T_{\rm in}^{-3.5}$. This in contrast to what is found for many BHBs (Fig.~\ref{fig2}) and might indicate that
the standard accretion disc theory is not a proper interpretation to the X-ray spectra from ULXs. This implies that the hypothesis on which
the IMBH idea is relying (i.e. standard accretion disc theory and the presence of a cold disc) are not valid and
it might indicate that these ULXs are not IMBHs as a ``class''. In the following we develop this idea into more detail.

We have seen above that the standard model for sub-critically accreting BHs \citep{shakura73} predicts the relation ${\rm L}{\propto}{\rm T}_{\rm max}^4$. \citet{poutanen07} develop a model
based on super-critical accretion. It is a model taking into account geometrical effects and where those related to advection are not fully taken into account. In spite of its simplicity it makes the 
relationship between observable amounts (e.g. temperature and luminosity from the spectra) and those we want to derive (i.e. the mass of the compact object) an easy task. Nevertheless, in the following Sec. we will
describe a fully consistent model based on the slim-disc calculations \citep{abramowicz88,sadowski11,bursa18}.

\begin{figure}[!t]
  \includegraphics[angle=0,clip,width=\columnwidth]{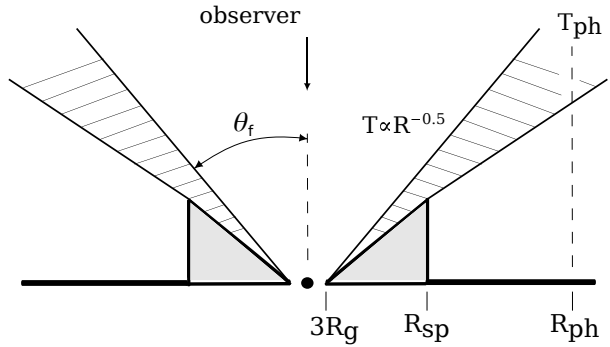}
  \caption{The model of a supercritical disc with a wind funnel. The figure shows the thin disc ($R > R_{\rm sp}$), the  supercritical disc ($R \leq R_{\rm sp}$), and 
the wind funnel constrained by the radius of the photosphere $R_{\rm ph}/\sin{\theta_f}$. From \citet{vinokurov13}.}
  \label{fig3}
\end{figure}

At super-Eddington accretion rates, three characteristic temperatures are identified: 

\begin{equation}
\label{eq:zones}
\begin{array}{ll}
\mbox{the maximal color disc temperature:} & \\
{\rm T}_{\rm c,max}={\rm f}_{\rm c}{\rm T}_{\rm max}\approx 1.6 {\rm f}_{\rm c}{\rm m}^{-1/4}\, \mbox{keV}, \\
\mbox{the color temperature at the spherization radius:} & \\
{\rm T}_{\rm c,sph} \approx 1.5 {\rm f}_{\rm c} {\rm  m}^{-1/4} \dot{m}_{0}^{-1/2}\, \mbox{keV}, \\
\mbox{the outer photosphere temperature given by:} &  \\
{\rm T}_{\rm ph} \approx m^{-1/4} \dot{m}_{0}^{-3/4}  \, \mbox{keV} .  
\end{array}
\end{equation}

\noindent where $\dot{m}_{0}= \dot{\rm M} / \dot{\rm M}_{\rm Edd}$ and ${\rm f}_{\rm c}$ is the correction factor to ${\rm T}_{\rm max}$. They are all marked at Fig.~\ref{fig3}.

A soft, $\sim$0.2 keV component may correspond to the spherization temperature implying the accretion rate $\dot{m}_{o}= m^{-1/2}(1.5f_{c}/{\rm T}_{\rm c,sph} [\mbox{keV}])^2$$ \approx$30--40 onto a
stellar mass, 10--20 ${\rm M}_{\odot}$, BH. The observed higher luminosities can result from the geometrical beaming. The absolute maximum apparent luminosity using this model (taking all the effects 
of advection and beaming of the flow) is about $10^{41}\,{\rm erg}{\rm s}^{-1}$ for a 20\,${\rm M}_{\odot}$ BH.

\subsection{The geometry of the disc}

A face-on observer would see the emission from three separate zones defined by the three characteristic radii (we refer to \citealt{poutanen07} for more details):

\begin{equation}
\label{eq:zones}
\begin{array}{ll}
r<{\rm r}_{\rm ph,in}, & \mbox{zone A} , \\
{\rm r}_{\rm ph,in} <r< {\rm r}_{\rm sph}, & \mbox{zone B} , \\
{\rm r}_{\rm sph} <r< {\rm r}_{\rm ph,out}, & \mbox{zone C} .
\end{array}
\end{equation}

The characteristic disc temperatures can be obtained from the Stefan-Boltzmann law (${\rm Q}_{\rm rad}(R)= {\sigma}_{\rm SB}{\rm T}^{4}(R)$) .

In zone A (i.e. outside from the photosphere), the wind is transparent (i.e. it is momentum-driven) and the radiation escapes unaffected by the outflow. 

In zone B, the wind is opaque and the energy generated in the disc is advected by the wind. The radiation escapes at a radius about twice the energy generation radius. This does not change the radial
dependence of the effective temperature $T\propto r^{-1/2}$ (i.e. slim-disc configuration),

The outer zone C emits about the Eddington luminosity which is produced mostly in the disc at radii ${\rm r}>{\rm r}_{\rm sph}$. The photon diffusion time here is smaller than the dynamical time, thus most
of the radiation escapes not far from the radius it is produced. This results in the effective temperature variation close to $r^{-3/4}$ (i.e. standard accretion regime; \citealt{shakura73}).

\subsection{The need of slim-disc models}

Typical models to fit the X-ray spectra from Ultra-luminous X-ray sources using XSPEC (e.g. {\tt DISKBB}, {\tt DISKPN}, {\tt KERRBB}, 
{\tt BHSPEC}, {\tt GRAD}; \citealt{arnaud96}\footnote{See the following web page for a list and description of the available models in XSPEC: \\https://heasarc.gsfc.nasa.gov/xanadu/xspec/manual/ }) are based 
on the thin disc model, which is not accurate for ${\rm L}>0.3\,{\rm L}_{\rm Edd}$. Indeed, such models
tend to give incorrect values for BH masses and for accretion rates (luminosities) above the Eddington limit, as shown in this Section.

\begin{figure}[!t]
  \includegraphics[angle=0,bb=191 310 420 481,clip,width=\columnwidth]{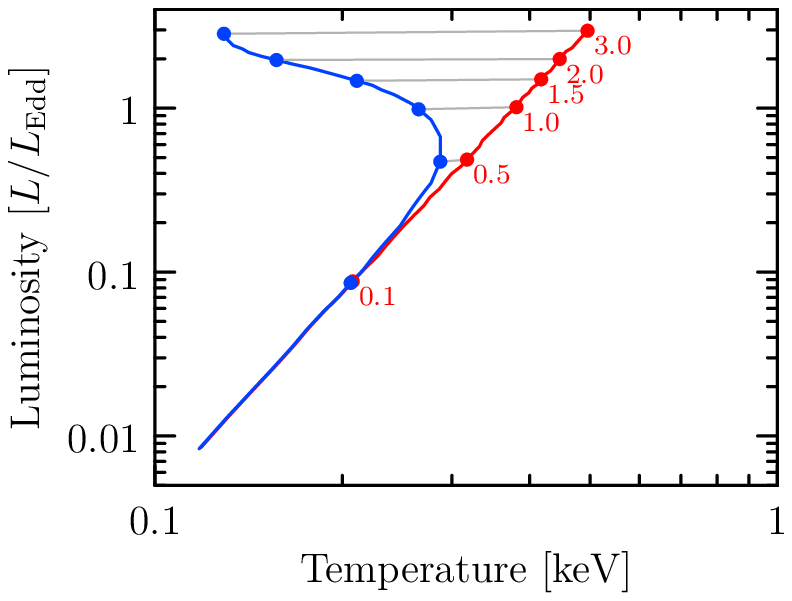}
  \includegraphics[angle=0,bb=191 310 420 481,clip,width=\columnwidth]{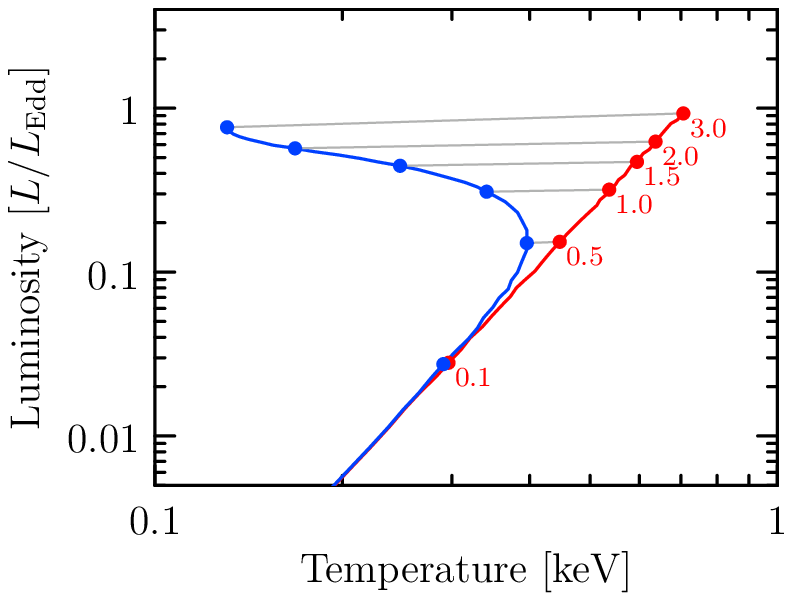}
  \caption{The luminosity-temperature relation for super-critically accreting BHs. The two panels correspond to two different inclination angles ($i=0^{\circ},70\,^{\circ}$, at top and bottom, respectively) of the
line of sight to the observer. Blue/red branches correspond to standard Novikov-Thorne (red) and slim (blue) accretion disc, with the labels indicating the accretion rate (in units of 
Eddington accretion rate). On the horizontal axis, the temperature [keV] denotes the best-fit temperature from both models (this temperature corresponds to the one giving the maximum emission flux in the calculated spectra). From \citet{bursa18}. }
  \label{fig4}
\end{figure}

The standard (thin) disc follows the ${\rm L}{\sim}{\rm T}^{4}$ relation (see Eq.~\ref{eq1}). Nevertheless, advection and obscuration effects cause 
significant deviations from that relation in the super-Eddington regime \citep{poutanen07}. These effects are taken into account in the so-called slim-disc 
models. This effect is strongly inclination dependent \citep{poutanen07} and the luminosity can stay around the Eddington limit even if the mass accretion rate
is much higher ($\dot{\rm M}{\gg}\dot{\rm M}_{\rm Edd}$). These facts have implications for the spectral modelling, e.g. getting lower inner disc temperatures
given a certain (high) luminosity (Fig.~\ref{fig2} and ~\ref{fig4}). This is in accordance to what has already been observed in the X-ray spectra from sample of representative 
ULXs \citep{poutanen07}.

Instead of using the classical models based on the thin disc theory here we present the results obtained by using the {\tt SLIMULX} 
model~\footnote{ http://stronggravity.eu/results/models-and-data/ }. {\tt SLIMULX} \citep{bursa18} is an 
additive model for thermal continuum emission at high accretion rates to be used with the X-ray spectral-fitting tool XSPEC. The model provides spectral
distribution of black-body radiation that is supposed to be emitted from the surface of a slim accretion disc \citep{abramowicz88}. A brief description
and a first application of the model can be found in \citet{mcaballe17}.

The major improvement of the {\tt SLIMULX} model over other commonly used disc models is that it includes three effects that are dominant at high accretion 
rates, and are not present in the standard \citet{shakura73,novikov73} disc model:
    
\begin{itemize}
\item Radial advection of heat, which plays a substantial role at higher luminosities, is present and significantly modifies the flux of radiation emitted at a given radius in the inner disc region.
    
\item The inner edge for the disc radiation deviates from ISCO at high luminosity and can be considerably closer to the black hole due to the advective transport of heat generated by viscous processes.
    
\item The location of the effective photosphere differs significantly from the equatorial plane with growing luminosity, although the relative vertical disc thickness is not large (h/r ${\le}$ 1), not even for the near-Eddington 
luminosities. Accordingly, the ray-tracing computations are taken from the actual disc photosphere to the observer at infinity \citep{sadowski09}.
\end{itemize}

\begin{figure}[!t]
  \includegraphics[angle=0,bb=0 0 1342 992,clip,width=\columnwidth]{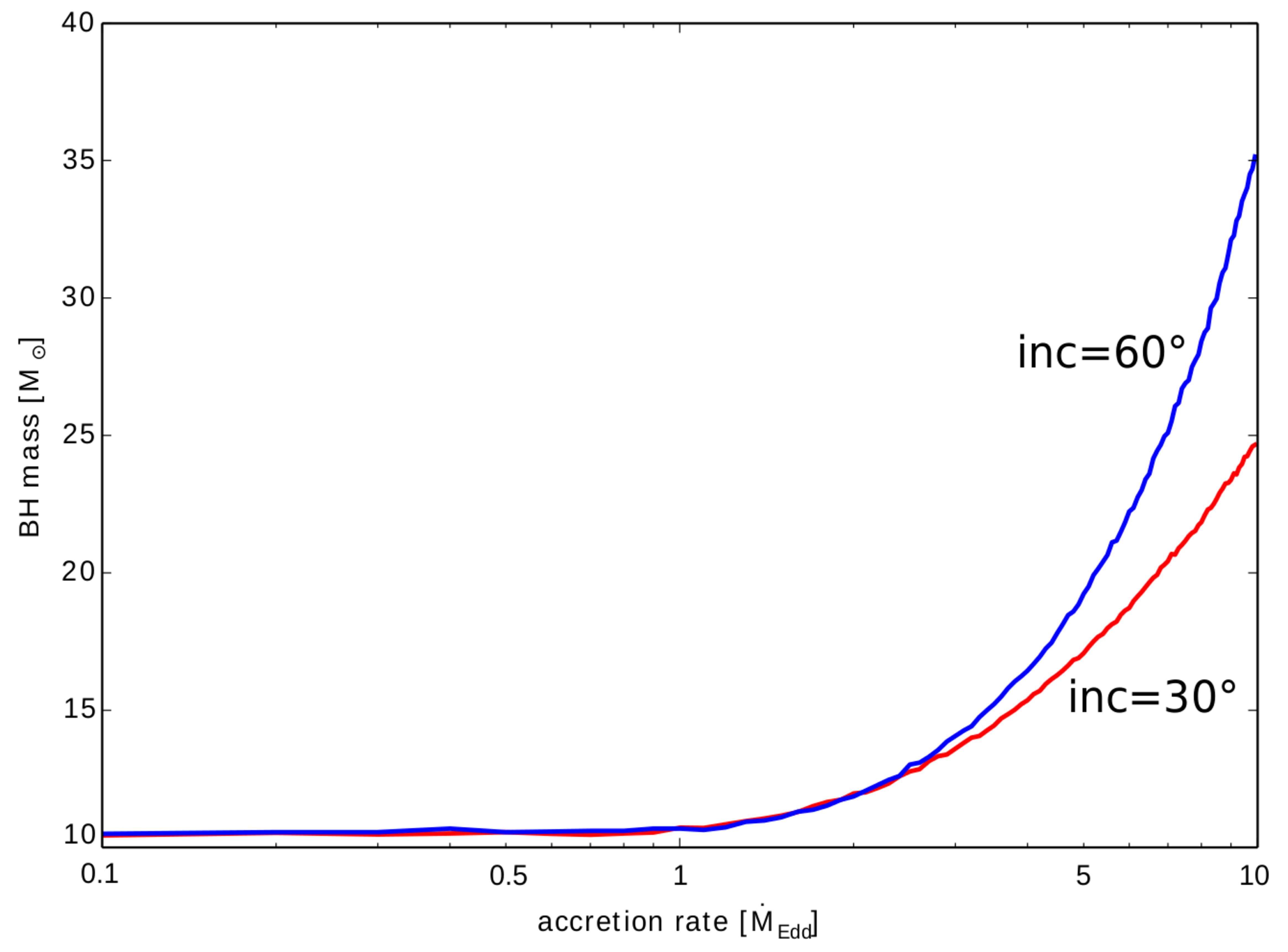}
  \caption{ Simulated {\tt SLIMULX} spectra are fitted with a thin disc model ({\tt DISKBB}; \citealt{mitsuda84,makishima86}) and the mass
is obtained from the fit (in units of ${\rm M}_{\odot}$). The horizontal axis is the accretion rate (in units of $\dot{\rm M}_{\rm Edd}$). From \citet{bursa18}. }
  \label{fig5}
\end{figure}

\citet{bursa18} fitted the simulated (i.e. faked) {\tt SLIMULX} spectra with a thin disc model ({\tt DISKBB}) and the mass was obtained from the fits (Fig.~\ref{fig5}). At low $\dot{\rm M}$, the
fit recovers the original mass (i.e. the one given by the slim-disc model), but at high mass accretion rate ($\dot{\rm M}{\ge}3\dot{\rm M}_{\rm Edd}$) the mass given
by the thin disc model is much higher (note that a mass of $10\,{\rm M}_{\odot}$ was assumed in the {\tt SLIMULX} simulated spectra). This exercise is representative of what 
is usually done when fitting the X-ray spectra from ULXs in the literature. It appears quite a bad approach to estimate the BH parameters when using thin disc models if the disc is strongly radiation pressure dominated.

So we should not use of Eqs.~\ref{eq1} and \ref{eq2} in the radiation pressure dominated (i.e. in the high accretion rate) regime.

In the {\tt SLIMULX} model the luminosity consists of the integration of all the radial flux elements (in the so-called corotating frame) multiplied by a factor that accounts for the proper transformation of the corotating frame
to the coordinate frame so that the given value is indeed the total luminosity of the disc ignoring geometrical GR effects. In this case the relationship between the mass and any measurable property of the disc
is troublesome. Even although the mass is a free parameter of the model, its direct relationship with the rest of quantities is complicated. It is generally assumed that the black hole mass is of a certain value
(e.g. 10\,${\rm M}_{\odot}$) when doing explicit calculations. Although relevant disc properties follow a mass scaling law or do not scale with mass, the results can not be directly rescaled and generalized to BH
systems of an arbitrary mass, because the output of radiative transfer calculation would be different and the predicted spectra also. The application to stellar mass BH systems is fine over an usually assumed mass range for
such systems (few to few tens of solar masses). Note that higher masses (e.g. IMBHs) can be used (and derived) with this model as well (see \citealt{straub14}).

\subsection{On the masses derived for the compact object using slim-disc models}

It has been suggested that ULXs appear very luminous due to a combination of moderately high mass (IMBHs), mild beaming and mild super-Eddington emission and that ULXs are
an inhomogeneous population composed of more than one class \citep{colbert99,fabbiano06}.

Initially, they were supposed to be the IMBHs originating from low-metallicity Population III stars \citep{madau01}. Nevertheless,
they are not spatially distributed throughout galaxies as it would be expected. On the other hand,
IMBHs may be produced in runaway mergers in the cores of young clusters \citep{portegies04}. In such cases, they usually
should stay within their clusters. Nevertheless, it has been found \citep{poutanen13} that all brightest X-ray sources in the Antennae galaxies are located
nearby the very young stellar clusters. NGC~5408 X--1 is also located nearby a young stellar association \citep{grise12}. These studies
concluded that these sources were ejected in the process of formation of stellar clusters in the dynamical few-body encounters and that the
majority of ULXs are massive X-ray binaries with the progenitor masses larger than $50\,{\rm M}_{\odot}$. Currently, it is thought that
only a handful of ULXs could be considered as potential IMBHs (ESO~243-49 HLX-1 between a few others; \citealt{farrell09,sutton12,mezcua17}).

Currently the most accepted idea is that the majority of ULXs are powered by accretion onto stellar-mass black holes $(<100\,{\rm M}_{\odot})$ at around or in
excess of the Eddington limit (e.g. \citealt{colbert99,fabrika01,king01,fabbiano06,poutanen07,liu13}).

Due their brightness, most of them are believed to be BHs. However, recently a new class of ULXs was discovered, through the detection of coherent pulsations: Ultra-luminous X-ray pulsars (ULPs). The presence of pulsations 
unambiguously identifies the compact objects as neutron stars, which are typically less massive than black holes. In ULPs the neutron star accretes matter from a companion star at inferred rates much 
higher than previously expected. Currently three of these systems are known: M82 X-2 \citep{bachetti14}, NGC 5907 ULX \citep{israel17a}, and NGC 7793 P13 \citep{israel17b}. Furthermore, a clear path 
forward to obtain a full sample of the ULP population is missing.

In this paper we have described all the models used to describe the data from ULXs, from assuming low accretion rates (and then high BH mass) to highly super-Eddington accretion rates (then low compact 
object mass). Applying the best models based on slim-disc configuration and including advection in the high accretion rate favour either low masses for the compact object 
(${\rm M}<30\,{\rm M}_{\odot}$, e.g. LMC X-3 and NGC5408 X-1 \citealt{straub11,mcaballe17}) or high masses (${\rm M}{\gtrsim}10^{3}\,{\rm M}_{\odot}$, ESO~243-49 HLX-1 \citealt{straub14}). Therefore BH masses in the range of
${\rm M}=100-10^{3}\,{\rm M}_{\odot}$ are hardly to find in the ULX population. Even though initially this population of ULXs was initially thought to contain a significant part of these IMBHs.

{\it Indeed, to our knowledge, there is no unambiguous detection of the electromagnetic counterpart of a BH with a mass in the range of
${\rm M}=80-10^{3}\,{\rm M}_{\odot}$}.

\subsection{X-ray Timing properties of ULXs}

Black Hole masses scale with the break frequency of their Power Density Spectrum (PDS; \citealt{mchardy06,kording07}). This relation holds over six orders of
magnitude in mass, i.e. from Black Hole Binaries (BHBs) to Super-Massive Black Holes (SMBHs).

\begin{figure}[!t]
  \includegraphics[angle=0,bb=18 180 558 648,clip,width=\columnwidth]{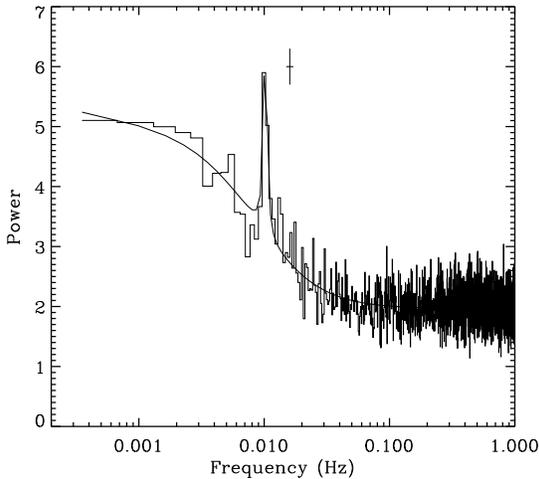}
  \caption{ Average Power Density Spectrum of NGC~5408 X--1. From \citealt{strohmayer09}.  }
  \label{fig6}
\end{figure}

The PDS and the energy spectra of NGC~5408 X--1 (Fig.~\ref{fig6}) and M~82 X--1 (for instance) are very similar to that of BHBs in the Steep Power-law state (i.e. one of the so-called
intermediate states). Nevertheless, the characteristic time-scales within the PDS are lower by a factor of a ${\approx}100$ and the X-ray luminosity is higher by a factor 
of a few ${\approx}10$, when compared to BHBs. This gives BH mass estimates for these ULXs of the order of ${\rm M}_{\rm BH}{\gtrsim}10^{3}-10^{4}\,{\rm M}_{\odot}$, from their
X-ray timing properties only, as explained in the following.

As described in the previous Sections, determining the mass from the BH in ULXs has been the goal of several studies. For example, in NGC~5408 X--1 there is still no consensus
on whether it is an IMBH or a stellar-mass BH. Previous estimates from the timing properties \citep{strohmayer09,dheeraj12} indicate a mass 
of ${\rm M}_{\rm BH}{\gtrsim}1\,000\,{\rm M}_{\odot}$, thus an IMBH, but others \citep{middleton11} indicate a much smaller mass of ${\rm M}_{\rm BH}{\le}100\,{\rm M}_{\odot}$, thus 
a stellar-mass BH \footnote{It has to be noted here that in low metallicity environments
BHs with masses up to $80-130\,{\rm M}_{\odot}$ can still be formed through direct stellar-collapse (\citealt{zampieri09,belczynski10}) and this is why we are referring to 
them as {\it stellar-mass} BHs.}. In the first case considered the accretion rate is sub-Eddington, whilst the latter case indicates (near or) super-Eddington accretion.

If the Quasi-Periodic Oscillation (QPO) detected in M~82 X--1 is a fundamental High Frequency QPO (HFQPO), then it appears at much lower frequency (${\approx}50$\,mHz) than those observed in BHBs (35--450\,Hz), by three orders of
magnitude. Scaling linearly with the mass of a stellar-mass BH ($10\,{\rm M}_{\odot}$) if these QPOs have the same origin, the frequencies we have found lead to a mass
of ${\rm M}_{\rm BH}{\approx}(10^4-10^{5})\,{\rm M}_{\odot}$ (taking into account the whole range of possible values of the spin of the BH) for the mass of the BH in M~82 X--1. Other
mass estimates, this time based on spectral-timing scaling relationships from systems of different mass \citep{titarchuk04}, have provided a mass estimate
of ${\approx}10^3\,{\rm M}_{\odot}$ \citep{fiorito04}. Nevertheless, caution is required in order to apply relationships based solely on the mass of the BH systems. As
already pointed out by several authors (\citealt{mchardy06} for AGNs; \citealt{soria07} and \citealt{casella08} for ULXs) the accretion rate is an important parameter that, together with
the mass of the BH, should be considered as the main drivers in these scaling relationships. This effect could lead to a much smaller mass for the BH (compact object). In addition, it remains to date unclear 
how timing properties and BH masses are related, and what are the properties of the accretion states in ULXs compared to those of Galactic BHBs \citep{mcaballe13a,atapin18}.

\section{Gravitational waves as a new window to the Universe}
\label{KK}

As reported by \citet{abbott16} the first gravitational-wave (GW) transient was identified in data recorded by the Advanced Laser Interferometer Gravitational-wave Observatory (LIGO) detectors 
on 2015 September 14. The event, initially designated G184098 and later given the name GW150914, is described in detail elsewhere. By prior arrangement, preliminary estimates of the time, significance, and 
sky location of the event were shared with 63 teams of observers covering radio, optical, near-infrared, X-ray, and gamma-ray wavelengths with ground- and space-based facilities. {\it As this event turned out to 
be a binary black hole merger, there is little expectation of a detectable electromagnetic (EM) signature}. GW150914 is consistent with the inspiral and merger of two BHs of masses $36^{+5}_{-4}$
and $29{\pm}4\,{\rm M}_{\odot}$, respectively, resulting in the formation of a final BH of mass $62{\pm}4\,{\rm M}_{\odot}$ at a distance of $410^{+160}_{-180}$\,Mpc.

As reported by \citet{abbott17} on 2017 August 17 a binary neutron star coalescence candidate (later designated GW170817) with merger time 12:41:04 UTC was observed through gravitational waves by 
the Advanced LIGO and Advanced Virgo detectors. The Fermi Gamma-ray Burst Monitor independently detected a gamma-ray burst (GRB 170817A) with a time delay of ${\approx}1.7$\,s with respect to the merger 
time. From the gravitational-wave signal, the source was initially localized to a sky region of 31\,deg$^{2}$ at a distance of $40_{-8}^{+8}$\,Mpc and with component masses consistent with neutron stars. The 
component masses were later measured to be in the range 0.86 to 2.26\,${\rm M}_{\odot}$. An extensive observing campaign was launched across the electromagnetic spectrum leading to the discovery of a bright optical 
transient (SSS17a, now with the IAU identification of AT 2017gfo) in NGC 4993 less than 11 hours after the merger. The follow-up observations support the hypothesis 
that GW170817 was produced by the merger of two neutron stars in NGC 4993 followed by a short gamma-ray burst (GRB 170817A) and a kilonova/macronova powered by the radioactive decay of r-process 
nuclei synthesized in the ejecta.

\begin{figure*}[!t]
  \includegraphics[angle=270,bb=0 0 595 842,clip,width=\textwidth]{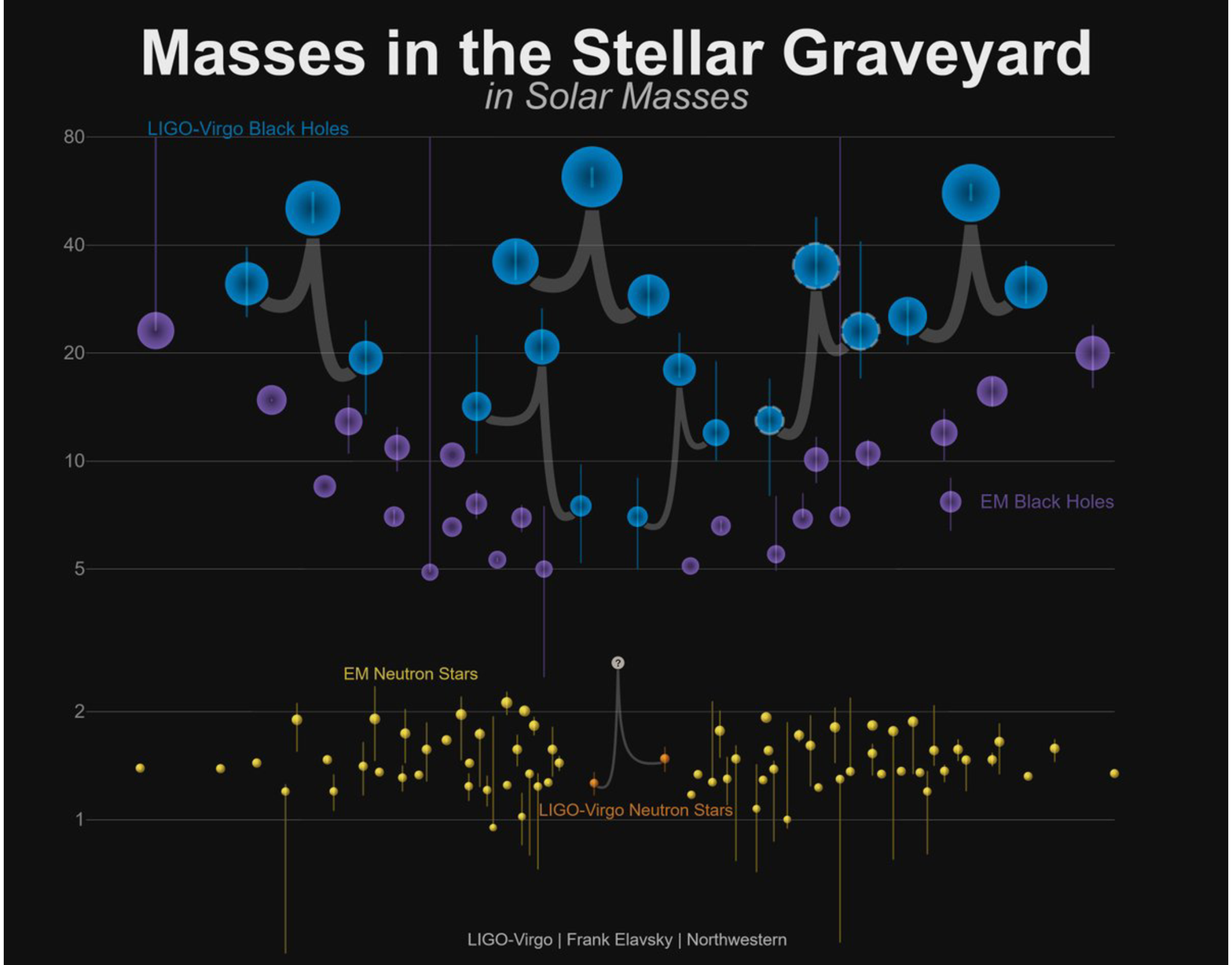}
  \caption{Interactive graphic featuring all of our BlackHoles (including GW170608), our recently announced
NeutronStars, and all the other compact objects known from electromagnetic measurements. From LIGO collaboration.}
  \label{fig7}
\end{figure*}

In Fig.~\ref{fig7} there is an interactive graphic featuring all the BH detected by LIGO (including GW170608), and the recently announced Neutron Stars, plus all the other compact objects known from 
electromagnetic measurements.

\subsection{Detection of the missing IMBHs}

Black holes, the ultra-compact remnants of very massive stars, are prime candidates for emitting detectable gravitational waves in this way. While current detection work focuses on 
commonly-expected binary systems of black holes with masses a few times that of the Sun, it is possible that there are detectable populations of black holes which have masses larger 
than this, i.e. hundreds or thousands times the mass of the Sun. These are dubbed IMBHs, as their masses lie in the currently-unobserved range between solar mass 
black holes and the supermassive black holes in the center of galaxies. The gravitational-wave interferometer network of LIGO and Virgo are able to measure ripples from the final 
moments of such intermediate mass binary black holes merging together.

Candidate IMBHs, nevertheless are still relatively rare (see our discussion above). Gravitational-wave detection of intermediate mass black hole binary mergers confirm their existence and 
provide information on the abundance of such systems in the Universe, as well as precisely measuring the component masses. The presence or absence of these signals will shed light on star 
formation in the early universe, provide important information about the dynamics and structures of globular clusters, and give clues about the formation of supermassive black holes.

Black holes are just that -- black. They are the corpses of dead stars, so massive and compact that not even light can escape them. They literally cannot be seen, and before LIGO came 
online, their existence could only be inferred by their gravitational effect on their neighbors or because of light (often X-rays) emitted by hot gas falling into the black hole.

But an isolated black hole is invisible. It interacts via gravity and, even then, it only emits gravitational radiation when it is moving. So detectors like LIGO or Virgo are the only 
way to see them. They are essentially black hole telescopes.

With even just a few observations of gravitational waves, the LIGO measurements are perplexing. Prior to 2016, it was thought that there were two classes of black holes: stellar-mass 
black holes, with masses no more than about 10 times that of our Sun, and massive, monstrous black holes at the center of galaxies with masses in the range of hundreds of thousands to billions of solar masses.

Black holes with masses in the range of ${\gtrsim}30-10^{3}\,{\rm M}_{\odot}$ were unexpected. And yet, that's just what LIGO (and now LIGO plus Virgo) have observed. It has
been shown that BHs in the mass range of $30-130\,{\rm M}_{\odot}$ can be formed through BH-BH mergers in low-metallicity environments (see recent simulations from \citealt{marchant16,marchant17}). They could 
justify the discovery of the recent IMBHs discovered by LIGO so far (${\le}130\,{\rm M}_{\odot}$; Fig.~\ref{fig7}). But observationally, IMBHs with masses in the range $100-10^3\,{\rm M}_{\odot}$ are 
extremely rare to find. They could form (part of) the unseen dark matter proposed to exist in the galactic haloes by cosmological studies (see e.g. \citealt{mediavilla17}, who tested for this possibility).

\section{Discussion and conclusions}
\label{KK}

Ultra-Luminous X-ray sources (ULXs) are accreting black holes for which their X-ray properties have been seen to be different to the case of stellar-mass black hole
binaries. For most of the cases their intrinsic energy spectra are well described by a cold accretion disc (thermal) plus a curved high-energy emission components. The
mass of the black hole (BH) derived from the thermal disc component is usually in the range of $100-10^{5}\,{\rm M}_{\odot}$, which have led to the idea that this might represent
strong evidence of the Intermediate Mass Black Holes (IMBH), proposed to exist by theoretical studies but with no firm detection (as a class) so far. But the mass estimation
depends on the accretion rate.

We have discussed the masses obtained by using different models considering different accretion rate regimes. Masses of the IMBH in the range of ${\rm M}=80-10^{3}\,{\rm M}_{\odot}$ are
hardly to obtain. Also, recent theoretical and observational developments are leading towards the idea that ULXs are instead stellar-mass compact objects accreting at an unusual super-Eddington 
regime instead, therefore favouring lower mass estimates for the compact objects. 

On the other hand, gravitational waves have been seen to be a useful tool for finding (some of) these IMBHs. We have given a brief overview about the recent advent of the discovery of
gravitational waves and their relationship with these so far elusive IMBHs.

\acknowledgments{MCG, MB and MD acknowledge support provided by the European Seventh Frame-work
Programme (FP7/2007-2013) under grant agreement n$^{\circ}$ 312789 and GA CR grant 18-00533S. SF 
acknowledges support by the Russian Science Foundation (N 14-50-00043) and the Russian Government 
Program of Competitive Growth of KFU. AJCT acknowledges support from the Spanish MINECO
Ministry project: AYA2015-71718-R. }

\end{document}